\documentclass[12pt]{article}
\usepackage{epsfig,graphicx}

\title{Nambu--Goldstone mesons in strong magnetic field}
\author{ V.D.Orlovsky and  Yu.A.Simonov,\\ Institute of Theoretical and Experimental
Physics\\ 117218, Moscow, B.Cheremushkinskaya 25, Russia}
\date{}

\newcommand{\be}{\begin{equation}}
\newcommand{\ee}{\end{equation}}

\def\la{\mathrel{\mathpalette\fun <}}
\def\ga{\mathrel{\mathpalette\fun >}}
\def\fun#1#2{\lower3.6pt\vbox{\baselineskip0pt\lineskip.9pt
\ialign{$\mathsurround=0pt#1\hfil ##\hfil$\crcr#2\crcr\sim\crcr}}}

\newcommand{{\SD}}{\rm SD}

\newcommand{\vex}{\mbox{\boldmath${\rm x}$}}

\newcommand{\vesig}{\mbox{\boldmath${\rm \sigma}$}}

\newcommand{\veP}{\mbox{\boldmath${\rm P}$}}
\newcommand{\vep}{\mbox{\boldmath${\rm p}$}}
\newcommand{\veq}{\mbox{\boldmath${\rm q}$}}

\newcommand{\vez}{\mbox{\boldmath${\rm z}$}}

\newcommand{\veL}{\mbox{\boldmath${\rm L}$}}

\newcommand{\veR}{\mbox{\boldmath${\rm R}$}}

\newcommand{\veta}{\mbox{\boldmath${\rm \eta}$}}
\newcommand{\veB}{\mbox{\boldmath${\rm B}$}}

\newcommand{\vepi}{\mbox{\boldmath${\rm \pi}$}}

\newcommand{{\Mc}}{\mathcal{M}}

\newcommand{\lan}{\langle}
\newcommand{\ran}{\rangle}

\begin{document}

\maketitle
\begin{abstract}
We study the $q\bar q$ structure embedded in chiral mesons in response  to
external magnetic fields (m.f.), using the chiral Lagrangian with $q\bar q$
degrees of freedom  derived earlier. We show that GMOR relations hold true for
neutral chiral mesons, while they are violated for the charged ones for
$eB>\sigma =0.2$ GeV$^2$. The standard chiral  perturbation theory also fails
in this region. Masses of $\pi^+$ and $\pi^0$ mesons are calculated and
compared to lattice data.
\end{abstract}

\section{Introduction}
Chiral Lagrangians introduced to clarify the dynamics of Nambu--Goldstone
mesons  have created a new selfconsistent formalism \cite{1} prior to the
emergence of QCD.

One of the basic  conceptual relations in QCD is the relation of the purely
chiral particles  -- the Nambu--Goldstone (NG) mesons -- to all other QCD
states, which are  mostly nonchiral. It other words one can define this as a
connection of Nambu--Goldstone to ordinary states, which can be called pure
confinement or the flux-tube states. The models treating most states in the
phenomenological chiral-like Lagrangians are now numerous, but unfortunately
they do not clarify this chiral -- confinement connection.

In \cite{2,3,4,5} one of the authors suggested a way of derivation the chiral
Nambu--Goldstone spectrum  from the QCD Lagrangian, where the basic chiral
relations: chiral condensate, GMOR relation and expressions for $m^2_\pi,
f^2_\pi$ are derived in terms of confinement (flux-tube) spectrum in the PS
isovector channel. The latter is calculated from the Relativistic Hamiltonian
in the framework of Field-Correlator Method  (FCM) \cite{6} in terms of the
basic input: current quark masses $m_i$, string tension $\sigma$ and
$\alpha_s$, and the FCM provides  a good description of the QCD spectrum in all
channels and for all masses $m_i$, except for NG mesons: $\pi, K, \eta$.

The connection of NG and flux-tube mesons \cite{2,3,4,5} described above, which
may be called the chiral-confinement relations (CCR),  allows to express NG
meson masses,   wave functions and quark decay constants in  terms of the same
basic input and in this way completes the theory. One should note, that in the
CCR one calculates not only ground states, but also excited NG states
\cite{4,5} and, moreover, one can study how chiral properties fade away with
growing quark current masses $m_i$ \cite{7}.

Recently a wide interest has occurred in the literature in the effects, which
can be produced in hadron dynamics due to strong external magnetic fields
(m.f.) \cite{8}. In particular, strong m.f. are expected in neutron stars
\cite{9}, early universe \cite{10}, heavy ion collisions \cite{11} and possibly
m.f. can produce strong reconstruction of the  vacuum \cite{12}.

From the theoretical point of view, strong m.f. play the role of crucial test
of the dynamics used in the model. For the QCD as a strong interaction theory
one must use the relativistic dynamical formalism, incorporating confinement
and perturbative gluon exchanges, producing all effects of strong decays. This
is naturally imbedded in the   FCM formalism, based on the QCD path integral,
where one derives the relativistic Hamiltonian (RH) for the $q\bar q, 3q$ etc.
states.

The inclusion of m.f. is done automatically in the RH, and the first results
for the masses were already obtained in \cite{13}, while the important role of
color Coulomb interaction  in strong m.f. was  studied in \cite{14}, and
magnetic moments of mesons in \cite{15}.

Of special interest is the influence of m.f. on chiral dynamics, and in this
way one can check that the CCR sustain their reliability in  the presence of
m.f. \cite{16}.
 On the lattice several analysis were done \cite{17, 18, 19} on chiral dynamics
 in m.f., e.g. the dependence of $\pi^+$ mass and chiral condensate on m.f. was
 done in unquenched QCD with physical pion mass \cite{20}. These results were
 compared with the CCR prediction for the $\lan \bar u u\ran$ and $\lan \bar d
 d\ran$ dependence on m.f. and a good agreement was found in  \cite{16}.

 On thee other hand, this dependence found on the lattice was  compared in
 \cite{20} with the what one expects from the chiral theory, and a strong
 disagreement was found for $eB>0.2$ GeV$^2$. This implied that the standard
 chiral theory \cite{4}, which lacks quark degrees of freedom, is unable to be
 a good working tool for distances less than 0.5 fm.

 In this paper we study the  m.f. dependence of the NG spectrum, which
 follows from CCR. The latter expresses  the NG masses through nonchiral PS
 isovector states, and we approximate several lowest states, which give the
 dominant contribution to CCR, in their m.f. dependence and obtain NG masses.

One of the most important result of this paper is the violation of GMOR
relations and  of the standard chiral formalism in the m.f. There appear
additional terms in the NG Lagrangian, proportional to m.f., which disclose the
internal quark-antiquark structure of NG mesons, not accounted for in the
standard chiral formalism of \cite{1}. As a result the  dependence of NG meson
mass on m.f. contains new terms, which do not vanish in the chiral limit $m_q
=0$. This behavior is supported by lattice data \cite{20} for the behavior of
the $\pi^+$ mass in m.f.

 In what follows we introduce first in section 2 RH for neutral and charged
 $q\bar q$ states in m.f., in section 3 we write down the basic equations of
 CCR and generalize them to the case of nonzero m.f. Section 4 is devoted to
 the calculation of NG states and  in section 5 results are compared  with chiral  perturbation
 theory and lattice data.

\section{Relativistic Hamiltonian for mesons in strong magnetic field }
 The RH for  the $q \bar q $ system in m.f. was derived recently in \cite{13, 13*} from the
 path integral in QCD and we follow these notations and definitions.

 \be H=H_0 + H_\sigma +W,\label{1}\ee
 where
\be H_0 =\sum^2_{i=1} \frac{\left(\vep^{(i)} - \frac{e_i}{2} (\veB\times
\vez^{(i)})\right)^2+ m^2_i +\omega^2_i}{2\omega_i},\label{2}\ee

 \be H_\sigma
=- \frac{e_1\vesig_1\veB}{2\omega_1}
-\frac{e_2\vesig_2\veB}{2\omega_2},\label{3}\ee

\be W=V_{\rm conf}+V_{\rm Coul}+V_{SD}+\Delta M_{ss}. \label{4}\ee One defines
$\omega_i \to \omega_i^{(0)}$ from the extremum values of  eigenvalues of the
operator $\bar H$

\be H_0+H_\sigma +V_{conf}  = \bar H;~~  \bar H\Psi=M_n^{(0)}\Psi,\label{5 }\ee

 \be \left.\frac {\partial
M_n^{(0)}(\omega_1\omega_2)}{\partial\omega_1}\right|_{\omega_i=\omega_i^{(0)}}=0,~~
i=1,2.\label{6 }\ee We now treat $H_0$ and try to separate c.m. and relative
motion

 \be \veR =
\frac{\omega_1\vez^{(1)}+\omega_2\vez^{(2)}}{\omega_1+\omega_2},~~\veta=
\vez^{(1)}-\vez^{(2)},\label{7}\ee

For two-body systems $q\bar q$ the c.m. and relative motion can be separated in
two cases:

 a)$ e_1+e_2=0,$ neutral case:

 b) $e_1=e_2, ~m_1 = m_2, ~ \omega_1= \omega_2$.

 We shall consider both cases below.

 In case a) one introduces the so-called ``phase factor'',

   \be \Psi(\veta,\veR)=\exp(i\Gamma) \varphi (\veta,
\veR),\label{8}\ee

\be \Gamma = \mathbf{P}\mathbf{R} - \frac{\bar e}{2} (\veB \times \veta) \veR,
~~ \bar e = \frac{e_1-e_2}{2},\label{9}\ee and defines  a new operator   $H'_0$
from the relation  \be H_0\Psi=\exp (i\Gamma) H'_0\varphi,\label{10}\ee

\be H'_0 = \frac{1}{2\tilde \omega} \left( - \frac{\partial^2}{\partial
     \veta^2}+\frac{e^2}{4} (\veB\times \veta)^2\right) +\sum^2_{i=1}\frac{m^2_i+\omega^2_i}{2\omega_i} ,\label{11}\ee

At this point it is convenient to replace linear confinement by the quadratic
one, with adjustable coefficient $\gamma$, which yields a deviation  $\la  5\%$
in resulting masses,
 \be V_{\rm conf} = \sigma \eta  \to \tilde V_{\rm conf}
= \frac{\sigma}{2} \left( \frac{\eta^2}{\gamma} + \gamma\right)\label{12}\ee
and now the average mass $M_n$ is the eigenvalue of the operator $\bar H'$ \be
\bar H' = H'_0 + H_\sigma + \tilde V_{\rm conf}; ~~ \bar H' \varphi= \bar M_n
\varphi,\label{13}\ee defines the extremal values of $(\omega_1, \omega_2,
\gamma) $

\be \left.\frac{\partial \bar M_n (\omega_1, \omega_2, \gamma)}{\partial
\omega_1}\right|_{\omega_1=\omega_1^{(0)}} =  \left.\frac{\partial \bar
M_n}{\partial \omega_2} \right|_{ (\omega_2=\omega_2^{(0)}}=
\left.\frac{\partial \bar M_n} {\partial \gamma}\right|_{\gamma=\gamma_0}
=0.\label{14}\ee

The resulting form of $\bar M_n^{(0)} = M_n (\omega_1^{(0)}, \omega_2^{(0)},
\gamma_0)$ defines the total mass of the meson, \be M_n = \bar M_n^{(0)} +
\Delta M_{\rm coul} + \Delta M_{SE} + \Delta M_{ss}.\label{15}\ee

The form of $\bar M_n$ (prior to stationary point insertions) is
 \be \bar M_n = \varepsilon_{n_\bot , n_z} +
\frac{m_1^2+\omega^2_1 - e\veB \vesig_1}{2\omega_1} +\frac{m_2^2+\omega^2_2 +
e\veB \vesig_2}{2\omega_2},\label{16}\ee where \be \varepsilon_{n_\bot, n_z} =
\frac{1}{2\tilde \omega} \left[ \sqrt{ e^2 B^2 + \frac{4\sigma\tilde
\omega}{\gamma}} (2n_\bot +1)+ \sqrt{\frac{4\sigma \tilde
\omega}{\gamma}}\left(n_z + \frac12\right)\right] + \frac{\gamma \sigma}{2}.
\label{17}\ee

The retaining  three terms in (\ref{15}) are defined as follows:

 1) $\Delta M_{\rm Coul} = \lan V_{\rm Coul}\ran$, where averaging is done with
  wave functions $\varphi_n$ defined in (\ref{13}), and $V_{\rm Coul}$ is the
  OGE interaction $V_{\rm Coul}(\veq) = \frac{-16 \pi\alpha_s}{3\veq^2}$, and at large
  m.f. $
  \veq^2$ is augmented by the $(q\bar q)$ loop contribution, see details in \cite{14}.

  2),~3) $\Delta M_{SE}$ and $\Delta M_{ss}$ are given in \cite{13} and we rewrite those
  in the Appendix.

  We now turn to the case b), and  consider two-body system with equal charges
  and masses. It is clear, that relativistic charged pions and kaons contain
  charges $e_1= \frac23 e, e_2=\frac13 e$, in contrast to the case b), however
  the main new feature in the case b)  is the contribution of the c.m. motion
  in m.f. to the total mass and this is pertinent also to the  realistic case,
  the difference between the case b) and the realistic case can possibly be
  treated in a perturbative manner. As it is clear from (\ref{8}), (\ref{9}),
  the phase factor  $\Gamma$ is not  necessary in the case b), and one obtains the
  Hamiltonian as in Eq.(\ref{1}), we also put below $\omega_1 =
  \omega_2=\omega, e_1=e_2=e.$

$$ H= \frac{P^2}{4\omega} - \frac{e(\veP(\veB \times \veR))}{2\omega} +
\frac{e^2}{4\omega} (\veB\times \veR)^2 + \frac{\pi^2}{\omega} + \frac{e^2}{16
\omega} (\veB\times \veta)^2+$$\be +  \frac{2m^2 + 2 \omega^2 - e (\vesig_1 +
\vesig_2) \veB}{2\omega} + \frac{\sigma}{2} \left(\frac{\eta^2}{\gamma} +
\gamma\right) + V_{\rm coul} + \Delta W.\label{18}\ee Solution of (\ref{18}),
treating $V_{\rm Coul}$ and $\Delta W$ as a perturbation, immediately yields

$$  M= \frac{m^2+\omega^2}{\omega} + \lan V_{\rm coul} \ran + \lan V_{ss} \ran
+ \lan \Delta M_{SE}\ran+$$\be + \frac{eB}{2\omega} (2 N_\bot +1) +
\sqrt{\left( \frac{eB}{2\omega}\right)^2 + \frac{2\sigma}{\gamma_0\omega}}
(2n_\bot +1) + (n_\parallel + \frac12) \sqrt{\frac{2\sigma}{\gamma_0 \omega}}-
\frac{e(\vesig_1+\vesig_2)\veB}{2\omega} + \frac{\gamma_0\sigma}{2},
\label{19}\ee where
$$ \gamma_0 = \beta_0 (B, \omega) \left( \frac{\sigma \omega}{2}\right)^{-1/3}$$
\be \beta_0^{3/2} (B,\omega) = \frac12 + \frac{1}{\sqrt{1+ \beta
(eB)^2{(4\sigma \omega)^{4/3}}}}\label{20}\ee

Finally, $\left. \frac{\partial M(\omega)}{\partial \omega}
\right|_{\omega=\omega_0} =0, ~~ \omega_0 (B) = a (B) \sqrt{\sigma}$.

For the lowest states and $eB\gg \sigma $ \be M^{ee}_0 = \omega + \frac{eB +
\sqrt{(eB)^2 + \bar c^2\sigma^2}-(\vesig_1+ \vesig_2) eB}{2\omega} +\textrm{const} \geq
0.\label{21}\ee

To be compared with the neutral case (Eq. (16) of our work) \be M_0^{e,-e} =
\omega+ \frac{1}{\omega} \sqrt{e^2 B^2 + \bar c^2 \sigma^2} - \frac{e\veB
(\vesig_1 - \vesig_2)}{2\omega}+... \geq0\label{22}\ee

In both cases no collapse due to spins.

One can see in (\ref{21}), that for the charged PS meson $\sigma_{1z} +
\sigma_{2z} =0$ and $M_0 (eB \to \infty)\approx 2\sqrt{eB} $, while for the
neutral case, Eq. (\ref{22}) for $\sigma_{2z} = - \sigma_{1z} =-1$ yields $M_0
(eB \to \infty) \to {\rm const}$.

However, for $\sigma_{1z} \neq \sigma_{2z}$ the stationary values of $\omega_1$
and $\omega_2$ can be different for large $eB$, and having our results in
(\ref{20}), (\ref{21}) as a first approximation, we now turn to the case $e_1
=\frac23 e, e_2 =\frac13 e, e>0$, and introducing the ``phase factor'' as in
(\ref{9}), with $\bar e= \frac{e_1-e_2}{2} = \frac{e}{6}$, one obtains the
Hamiltonian

$$ H'_0=\frac{\veP^2}{2(\omega_1+\omega_2)}+\frac{(\omega_1+\omega_2)\Omega^2_R\veR^2_\bot}{2}
+\frac{\vepi^2}{2\tilde \omega}+ \frac{\tilde \omega
\Omega^2_\eta\veta^2_\bot}{2}+X_{LP}\veB\veL_P+$$

$$+X_{L_{\eta}}\veB\veL_\eta+X_1\veP(\veB\times\veta)+X_2(\veB\times \veR)\cdot(\veB\times\veta)+$$

\be+ X_3\vepi(\veB\times \veR)+\frac{m^2_1+\omega^2_1}{2\omega_1}+
\frac{m^2_2+\omega^2_2}{2\omega_2}.\label{23}\ee


Here $\Omega_R, \Omega_\eta$ are

\be\Omega^2_R= B^2\frac{(e_1+e_2)^2}{16\omega_1\omega_2}\label{24}\ee \be
\Omega^2_\eta=\frac{B^2}{2\tilde \omega(\omega_1+\omega_2)^2} \left[
\frac{(e_1\omega_2+\bar e\omega_1)^2}{2\omega_1}+\frac{(e_2\omega_1-\bar
e\omega_2)^2} {2\omega_2}\right].\label{25}\ee All coefficients $X_i (i=1,2,3)$
are given in the Appendix 1 of \cite{15}.

One can see in (\ref{23}) that the c.m. and relative coordinates can be
separated, provided the terms $X_1, X_2, X_3$  vanish, or else one can treat
them as a perturbative correction \be \Delta M_X = \lan  X_1 \veP (\veB\times
\veta) + X_2 (\veB \times \veR) (\veB \times \veta) + X_3 \vepi (\veB \times
\veR) \ran.\label{26}\ee
  Then one can write the total eigenvalue  $M_n^{(0)}$ of the  Hamiltonian
  $\bar H'$ in (\ref{13}) as \be M_n^{(0)} = M^{(0)} (\veP) + M^{(0)} (\vepi)+
  \Delta M_X + H_\sigma \label{27}\ee
  where
  \be M^{(0)} (\veP) = \frac{P^2_z}{2(\omega_1 +\omega_2)} + \Omega_R (2
  n_{R_\bot}+1) +X_{LP} \veL_P \veB,\label{28}\ee
  $M^{(0)}(\vepi)$ is the eigenvalue of the operator $H_\pi$,
  \be h_\pi = \frac{\vepi^2}{2\tilde\omega}+ \frac{\tilde \omega
  \Omega^\Omega_\eta \veta^2_\bot}{2} +X_{L_\eta} \veB \veL_\eta + V_{conf}
  +V_{OGE}.\label{29}\ee

Note, that we take in the zeroth approximation the $q\bar q$ state with $\veL_p
= \veL_\eta =0$, in which case $\Delta M_X$ vanishes in the first
approximation, and one has the following result for the lowest mass, as in
(\ref{15}), (\ref{16}), but with additional c.m. contribution $\Omega_R$.\be
M_n = \bar M_n^{(0)} + \Delta M_{\rm Coul} + \Delta M_{SE} + \Delta
M_{ss}\label{30}\ee \be \bar M_n = \Omega_R + \varepsilon^{(+)}_{n_\bot, n_z} +
\frac{m_1^2+\omega_1^2- \frac23 e \veB
\vesig_1}{2\omega_1}+\frac{m_2^2+\omega_2^2- \frac13 e \veB
\vesig_2}{2\omega_2}\label{31}\ee where \be \varepsilon^{(+)}_{n_\bot, n_z}=
\sqrt{\Omega^2_\eta + \frac{\sigma}{\gamma\tilde \omega}} (2n_\bot +1) + \sqrt{
\frac{\sigma}{\gamma\tilde \omega} }(n_z + \frac12) + \frac{\gamma
\sigma}{2}\label{32}\ee and one can see, that at large $eB \gg \sigma, ~~\bar
M_n$ has the form (for $n_\bot =n_z =0)$ \be \bar M_n (eB \to \infty) =\Omega_R
+ \Omega_\eta + \frac{\omega_1+ \omega_2}{2} - \frac23 \frac{e\veB
\vesig_1}{2\omega_1} - \frac13 \frac{e\veB \vesig_2}{2\omega_2}\label{33}\ee
provided $\Delta M_{SE}$ and $\Delta M_{ss}$ grow  slower than $eB$.

\section{Derivation of the effective chiral lagrangian in m.f.}

We follow here \cite{3,4} to write first the effective lagrangian of the light
quark in the confining field of the antiquark   in m.f., starting with the
standard QCD partition function in Euclidean space-time \be Z= \int DA D\psi
D\psi^+ \exp [L_0 + L_1 + L_{int}]\label{1}\ee where \be L_0 = -\frac14 \int
d^4 x (F^a_{\mu\nu})^2,\label{b2}\ee

\be L_1 =-i \int~^f\psi^+(x) (\hat D+ m_f) ^f\psi(x) d^4x\label{b3}\ee  \be
L_{int} = \int~^f\psi^+(x)  g\hat A(x) ^f\psi(x) d^4x\label{b4}\ee and  \be
\hat D = \gamma_\mu (\partial_\mu - ie_f A_\mu^{(e)} (x)), ~~ A_{\mu}
^{(e)}(x)=\frac12 [\vex\times \veB].\label{b5}\ee

Note, that in m.f. $\hat D$ can be considered as a diagonal $2\times 2$ matrix
in the flavor space, with  $e_f=e_u$ or $e_d$.

Averaging $Z$ over vacuum gluonic field and keeping only lowest (bilocal)
correlators of color fields $D_{\mu\nu,\lambda\sigma} (x,y) =\frac{1}{N_c} tr
\lan F_{\mu\nu}(x) \Phi(x,y) F_{\lambda\sigma} (y) \Phi(y, x)\ran$, one finds
as in \cite{2,3,4,5}, \be \lan Z\ran_A = \int D\psi D\psi^+ \exp (L_1 +
L_{eff}^{(4)}), \label{b6}\ee where \be L_{eff}^{(4)} =\frac{1}{2N_c} \int d^4
x d^4 y ~^f\psi^+_{a\alpha} (x)^f\psi_{b\beta}(x)^g\psi_{b\gamma}
(y)^g\psi_{a\delta} (y)J_{\alpha\beta, \gamma\delta}(x,y)\label{b7}\ee and \be
J_{\alpha\beta, \gamma\delta} (x,y) = (\gamma_\mu)_{\alpha\beta}
(\gamma_\nu)_{\gamma\delta} J_{\mu\nu} (x,y)\label{b8}\ee \be J_{\mu\nu} (x,y)
= g^2\int^x_C du_\alpha \int^y_C dv_\beta D_{\alpha\mu,\beta\nu}
(u,v)\label{b9}\ee

Here indices $f,g$ refer to flavor, $a,b$ to color and $\alpha,\beta, \mu,\nu$
to Lorentz indices.  Eq.  (\ref{b9}) implies that some contour gauge is used
for simplicity, but the final result is gauge invariant and the most important
 property of  $J_{\mu\nu}(x,y)$ is that it is proportional to the distance of
 the average point $\left(\frac{x+y}{2}\right)$ to the contour $C$ (linear
 confinement), and the effective distance between $x$ and $y$ (nonlocality) is
 of the order of the vacuum correlation length $\lambda \approx 0.1$ fm.

 In the large $N_c$ limit the four-quark expression in $L^{(4)}_{eff}$ can be
 replaced by the quadratic one, using the limit
 \be ^f\psi_{b\beta} (x) ^g\psi_{b\gamma} (y) \to \delta_{fg}
 N_c~^fS_{\beta\gamma}(x,y), \label{b10}\ee
 where $^fS_{\beta\gamma} (x,y)$ is the quark propagator.

As  a result one obtains the form \be L_{eff}^{(4)} \to - i \int
d^4xd^4y~^f\psi^+_{a\alpha} (x)^{(fg)} M_{\alpha\beta} (x,y)
~^g\psi_{a\beta}(y),\label{b11}\ee

\be ^{fg}M_{\alpha\delta} (x,y) =-iJ_{\mu\nu} (x,y) (\gamma_\mu~^{fg}S(x,y)
\gamma_\nu)_{\alpha\delta},\label{b12}\ee and the quark propagator satisfies
the equation \be (-i\hat D - im_f) ~^f S(x,y) - i \int~^{(fg)} M(x,z) ~^gS(z,y)
= \delta^{(4)}(x,y).\label{b13}\ee It is convenient to use the following
parametrization of $^fM(x,y)$ in terms of scalar functions, flavor singlet
$M_s(x,y)$ and flavor triplet $\phi_a(x,y) , a=1,2,3$,\be
M_{\alpha\beta}^{(fg)} x,y) =M_s (x,y) \exp (i\gamma_5 t^a \phi_a (x,y)
)^{(fg)}_{\alpha\beta}\equiv M_s (x,y) \hat U_{\alpha\beta}^{(fg)}
(x,y)\label{b14}\ee As a result the effective Lagrangian assumes the form \be
L_{\phi} = \int d^4x d^4y \left\{ ^f\psi^+_{a\alpha} (x) [(i\hat D +
im_f)_{\alpha\beta} \delta^{(4)}(x-y) \delta_{fg}+ iM_s\hat
U_{\alpha\beta}^{(fg)} (x,y)] ~^g\psi_{a\beta} (y)\right\},\label{b15}\ee and
the partition function can be written as \be \lan Z\ran_A=\int  D\psi D\psi^+ D
M_s D\phi_a \exp L_\phi.\label{b16}\ee

Integrating over $D\psi D\psi^+$ one obtains the effective chiral Lagrangian
(ECL) $L_{ECL},$ \be \lan Z\ran_A = \int DM_s D\phi_a \exp
L_{ECL},\label{b17}\ee where

\be L_{ECL} =N_c \textrm{tr} log [(i\hat D+ im_f)\hat 1 + iM_s\hat U].\label{b18}\ee

Finally,the ECL at the stationary point in the integral (\ref{b17}) is defined
by conditions $\frac{\delta L_{ECL}}{\delta M_S}= \frac{\delta L_{ECL}}{\delta
\phi_a}=0,$which yields \be iM_s^{(0)} (x,y) = (\gamma_\mu S^{(0)}
\gamma_\nu)J_{\mu\nu} (x,y), ~~ \phi_a^{(0)}=0,\label{b19}\ee and $$
S^{(0)}=S_\phi (\phi_a=0),~~ S_\phi =- [(i\hat D+im_f) \hat 1 + i M_s^{(0)}\hat
U]^{-1}.$$

Insertion of (\ref{b19}) in (\ref{b18}) yields the effective action for
pseudoscalar fields $\phi_a$

\be L_{ECL} \to -W (\phi)=N_c \textrm{tr} \log [ (i\hat D+ im_f)\hat 1 + iM_s^{(0)}\hat
U].\label{b20}\ee Our final step here is the  local limit of  $J_{\mu\nu}(x,y)$
and $M_s^{(0)}(x,y)$ proved in \cite{2,3,4,5}, which yields \be \phi_a (x,y)
\to \phi_a(x), ~~ M_s^{(0)}(x,y)\to M_s^{(0} (x)
\delta^{(4)}(x-y)\label{b21}\ee Expanding $W(\phi)$ in powers of $\phi_a$ and
keeping quadratic terms, one has \be W^{(2)}(\phi)=\frac12 \int \frac{d^4k
d^4k'}{(2\pi)^4(2\pi)^4} \phi_a^+ (k) N(k,k') \phi_a(k'), \label{b22}\ee where
$$\hat N(k,k')=\frac{N_c}{2}\int
dx e^{i(k+k')x} tr (t_a\Lambda M_st_a)_{xx}+
$$

$$ \frac{N_c}{2}\int d^4(x-y)d^4 \left( \frac{x+y}{2}\right) \exp\left[ i
\frac{(k+k')}{2}(x+y)+\frac{i}{2}(k-k')(y-x)\right]\times$$
 \be \times
\textrm{tr}[\Lambda (x,y)t_a M_s (y)\bar \Lambda(y,x) t_aM_s(x)],\label{bb22}\ee with
the definitions $M_s \equiv M_s^{(0)},$ \be \Lambda=(\hat D +m+M_s)^{-1},~~
\bar \Lambda=(\hat D-m-M_s)^{-1}.\label{b23}\ee

It is important at this point to make explicit the dependence of $\Lambda$ and
$\bar \Lambda$ on charges. We shall consider below the cases of neutral and
charged NG mesons and their  treatment will be different, since charged NG
mesons contain additional c.m. term in m.f. We start with the neutral case and
define in (\ref{b22}), (\ref{bb22}) $a=3$ and $\Lambda_+ = (\hat D_+ +
m+M_s)^{-1}, \bar \Lambda_+ =(\hat D_+ - m-M_s)^{-1}, \hat D_{+/-} = \hat
\partial \mp i e_q \hat A^{(e)}, $ and the same  for $\Lambda_-, \bar \Lambda_-$.

Using  translation invariance of traces in (\ref{b22}) one can rewrite it for
neutral NG mesons as \be W^{(2)}(\phi)=\frac{N_c}{2}\int
\phi_3(k)\phi_3(-k)\bar N_{33}(k)\frac{d^4 k}{(2\pi)^4},\label{b24}\ee where
\be \bar N_{33}  (k) =\frac12 \textrm{tr} \{ (\frac{\Lambda_++\Lambda_-}{2}
M_s)_0+\frac12\int d^4ze^{ikz}\Lambda_+(0,z) M_s(0) \bar \Lambda_- ( z
,0)M_s(0)+\label{b25}\ee $$ +  \frac12 \int d^4ze^{ikz} \Lambda_- (0,z) M_s (0)
\bar \Lambda_+ (z,0) M_s (0)\}.$$



At  this point  it is important to make clear, how the GMOR relations occur
from the effective Lagrangian (\ref{b22}) from the expression for $\bar N (0)$
in the case, when m.f. is absent, and how they are violated by m.f.

In the case of no m.f. one can write \be \bar N(0) = \frac12 \textrm{tr} \left\{ \Lambda
M_s + \Lambda M_s \bar \Lambda M_s\right\}=\frac12 \textrm{tr} \left\{ \Lambda M_s \bar
\Lambda (\hat \partial- m) \right\} =\label{60} \ee$$= \frac{m}{4} \textrm{tr} (\Lambda-
\bar \Lambda) = \frac12 m \textrm{tr} \Lambda + O(m^2),$$
where we have used identity $\bar \Lambda (\hat \partial-m-M_s)=1$ in the first
step, vanishing of the vector part of the expression in the second step, and
another identity $M_s = \Lambda^{-1}- \bar \Lambda^{-1}-m $ in the last step.

Since  $ \bar N (0) = \frac{m^2_\pi f^2_\pi}{4N_c}$, one obtains in (\ref{60})
the GMOR relation, as shown in \cite{3,4}.

Another situation occurs in the case of m.f. Indeed, Eq. (\ref{60}) in this
case acquires the form \be \bar N_{33} (0) = \frac14 \textrm{tr} \{ (\Lambda_+ +
\Lambda_-) M_s + \Lambda_+ M_s \bar \Lambda_- M_s + \Lambda_- M_s \bar
\Lambda_+ M_s\},\label{61}\ee and one can rewrite this expression as \be \bar
N_{33} (0) = \frac12 \textrm{tr} \left\{ -\frac{m}{2} (\Lambda_+ M_s\bar  \Lambda_-+
\Lambda_- M_s \bar \Lambda_+) + \Delta \bar N_{33} (0)\right\} \label{62}\ee
where the new term is \be \Delta \bar N_{33} (0) = \frac{M_s (m+ M_s)}{2} \textrm{tr} [
G_- \hat D_- \hat D_+ G_+ - G_- \hat D_- G_+ \hat D_- + \label{63}\ee$$+\hat
D_- G_- G_+ \hat D_+ - G_- \hat D_+ G_+ \hat D_+],$$
and we have introduced quadratic Green's functions \be G_+ \equiv
\frac{1}{(m+M_s)^2- \hat D^2_+}, ~~ G_- = \frac{1}{(m+M_s)^2 - \hat
D^2_-}\label{64}\ee

In (\ref{63}) $\hat D_-, \hat D_+$ are acting at the vertices as follows e.g.
for the second term inside the \textrm{tr} sign,

$$ \int \hat D_-(x) G_+ (x,y) \hat D_- (y) G_-(y,x) d^4(x-y)\Rightarrow$$
\be \Rightarrow \int \hat D_- (x) \lan x|e^{-\hat H_{+-} T} |y\ran \hat D_- (y)
d^4(x-y)=\label{65]}\ee
$$ \int\lan x| [m - i(\hat p_- + 2e\hat A^{(e)}(x))] e^{-M_+ T} (m-i\hat p_-)
|y\ran d^4(x-y).$$

However $A^e(\vex =0)=0$, and therefore magnetic field  $\veB$ cannot act on
charges at the vertices $x,y$ but only can act via the magnetic moment terms,
which are the same in the denominators of all  four terms in (\ref{63}), but
these terms also appear in the products $\hat D_-\hat D_+$ and $\hat D_+\hat
D_-$ in the first and third term under the \textrm{tr} sign, namely

\be\hat D^2_+ = (D^+_\mu)^2 + e\vesig \veB, ~~ \hat D_+\hat D_- = D_\mu^+
D^-_\mu - e\vesig \veB,\label{66}\ee

$$\hat D^2_- = (D^-_\mu)^2 - e\vesig \veB, ~~ \hat D_-\hat D_+ = D_\mu^-  D^+_\mu + e\vesig \veB.$$

Therefore in the sum of these terms, $ \hat D_-\hat D_+  G_+G_-+ \hat D_+\hat
D_-G_-G_+$ one can take into account, that $G_+$ and $G_-$ commute in the
constant m.f. and therefore the sum due to (\ref{66}) vanishes. Thus we come to
the conclusion, that $\Delta \bar N_{33} (0)$ for neutral mesons vanishes, and
the GMOR relation survives with $\bar N_{33} (0)= \frac{m^2_\pi f^2_\pi}{4N_c}$
and $\bar N_{33}(0)$  is given in (\ref{62}) with $\Delta \bar N_{33} (0)=0$.

The calculation of the quark condensate in this case was done in \cite{16}.

\section{Masses of NG mesons in magnetic field}

We start with the mass of the neutral NG meson and as shown in the previous
section, one can use the GMOR relation with m.f. induced quark condensate and
$f_\pi$.

The GMOR relation with additional $O(m^2)$ correction, found in  \cite{7}, is
\be m^2_\pi f^2_\pi = \frac{\bar m M(0)}{M(0)+\bar m} |\lan \bar u u \ran +
 \lan \bar d d \ran|, ~~ \bar m = \frac{m_u + m_d}{2},\label{67a}\ee and
the quark condensate in m.f. was defined in \cite{16}, \be |\lan \bar q q\ran_i
| = N_c (M(0) + m_i ) \sum^\infty_{n=0} \left( \frac{\frac 12
|\psi_{n,i}^{(+-)} (0) |^2}{m_{n,i}^{(+-)}}+\frac{\frac12 |\psi_{n,i}^{(-+)}
(0) |^2}{m_{n,i}^{(-+)}}\right),\label{68a}\ee where $i=u,d,s$ and the
superscripts$(+-)$ and $(-+)$ refer to the quark antiquark spin projections on
m.f. $\veB$. In a similar way one can use the derivation of $f^2_\pi$, given in
\cite{3,4} to generalize it to $(+-)$ and $(-+)$ projections of the Green's
function, namely \be  f^2_\pi =  N_c M^2(0)  \sum^\infty_{n=0} \left(
\frac{\frac12 |\psi_{n,i}^{(+-)} (0) |^2}{(m_{n,i}^{(+-)})^3}+\frac{\frac12
|\psi_{n,i}^{(-+)} (0) |^2}{(m_{n,i}^{(-+)})^3}\right).\label{69a}\ee

Actually in (\ref{68a}), (\ref{69a}) the summation is over $n\equiv (n_3,
n_\bot)$ and while masses $m_{n_3, n_\bot}$ strongly grow with $n_\bot$ in
m.f., the sum over $n_3$ cut off due to factors $\exp (-m_n\lambda)$ in
(\ref{68a}) and $\exp (-m_n\lambda)(1+m_n\lambda)$ in (\ref{69a}), see
\cite{3,4} for details. As a result only few first terms contribute in
(\ref{68a}), (\ref{69a}), and as was argued in \cite{16} one can replace
$|\psi_{n,i} (0)|^2$ by \be |\psi^{(+-)}_{n_\bot= 0,n_3}
(0)|^2\cong\frac{\sqrt{\sigma} \sqrt{e^2_q
B^2+\sigma}}{(2\pi)^{3/2}},\label{70a}\ee

\be |\psi^{(-+)}_{n_\bot= 0,n_3} (0)|^2\cong(\sigma^2 c_{-+})^{3/4} \sqrt{1+
\left(\frac{e_qB}{\sigma}\right)^2\frac{1}{c_{-+}}},\label{71a}\ee and $c_{-+}
(B) = \left( 1+ \frac{8e_q B}{\sigma}\right)^{2/3}$

As it  is seen in (\ref{16}), (\ref{17}), (\ref{22}), the mass of the $(+-)$
state does not grow with $|e_qB|$, $m^{(+-)}_{n_\bot =0,n_3}
=O(\sqrt{\sigma})$, while  the mass of the $(-+)$ state grows as $2\sqrt{2|e_q
B|+\frac{\sigma}{4}},$ therefore we can neglect the sum over $(-+)$ states in
(\ref{69a}), and write  for $eB\gg \sigma$ \be f^2_\pi (B) =
\frac{N_cM^2(0)}{(\bar m^{(+-)})^2} \sum\frac{\frac12 |\psi_{n,i}^{(+-)} (0)
|^2}{(m_{n,i}^{(+-)})}.\label{72}\ee

On the other hand,  $|\lan \bar q q\ran_i|$ was estimated, using (\ref{68a}),
in \cite{16}  as \be |\lan \bar q q\ran_i (B) | =| \lan \bar q q\ran_i(0) |
\frac12 \left\{ \sqrt{1+\left(\frac{e_qB}{\sigma}\right)^2}+
\sqrt{1+\left(\frac{e_qB}{\sigma}\right)^2\frac{1}{c_{-+}}}\right \}
\label{73}\ee and finally the mass of $\pi^0$ at large m.f. $|eB|\gg \sigma$
can be written as \be m^2_\pi = \frac{\bar m}{M(0)} (\bar m^{(+-)})^2 \left\{
1+ A  \right\},~~~A=\left[ \frac{1+ \left(
\frac{e_qB}{\sigma}\right)^2\frac{1}{c_{-+}}}{1+ \left(
\frac{e_qB}{\sigma}\right)^2}\right]^{1/2}\label{74}\ee where $\bar m^{(+-)}$
is close to the lowest $(+-)$ mass with $n_\bot =n_3 =0$.


\begin{figure}
  \center{\includegraphics[height=7.0cm]{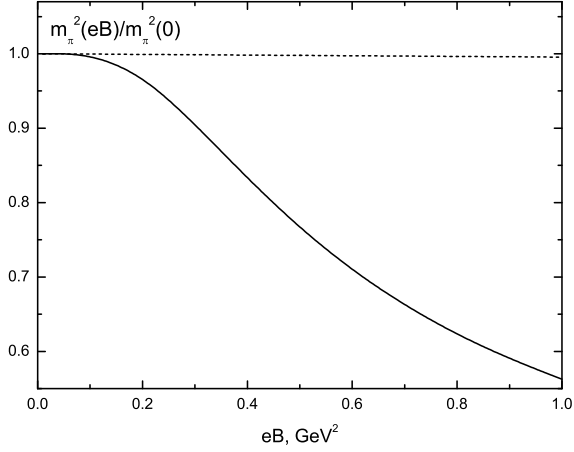}
    \caption{The normalized mass of $\pi^0$ meson as a function of magnetic field strength (solid line) in comparison with prediction of ChPT (\ref{90}) (broken line). \label{fig1}}}
\end{figure}

We can find the $\pi^0$ mass numerically, keeping for $|\lan \bar q q\ran|$ and $f_{\pi}^2$ the first few terms in sums over $n$ (\ref{68a}) and (\ref{69a}). The masses $m_{n,i}^{(+-)}$ and $m_{n,i}^{(-+)}$ are taken as eigenvalues (\ref{16}), (\ref{17}) with appropriate spin directions, while for the values of wave function we have the following expression:
\be
|\psi_{n_1,n_2,n_3}(0)|^2 = \frac{n_1!n_2!n_3!}{2^{n_1}(\frac{n_1}{2})!^22^{n_2}(\frac{n_2}{2})!^22^{n_3}(\frac{n_3}{2})!^2\pi^{3/2}r_\perp^2r_0},
\ee
if all $n_1, n_2, n_3$ are even, for odd $n_1, n_2$ or $n_3$ $|\psi_{n_1,n_2,n_3}(0)|^2=0$. The transverse and longitudinal scale parameters $r_\bot = \sqrt{\frac{2}{eB}} \left( 1+ \frac{4\sigma\tilde \omega}{\gamma e^2B^2}\right)^{-1/4},~ r_0 = \left( \frac{\gamma}{\sigma \tilde \omega}\right)^{1/4}$. The cut-off parameter $\lambda$ is taken to be about $1$ GeV$^{-1}$. The resulting normalized mass $\frac{m^2_{\pi^0}(eB)}{m^2_{\pi^0}(0)}$ is given in Fig. 1 in comparison with prediction of chiral perturbation theory (ChPT) (\ref{90}). This behavior is in agreement with lattice data for $\pi^0$ in \cite{21}.

We now turn to the case of charged mesons, e.g. $\pi^+$, and one can expect,
that, neglecting the internal structure of $\pi^+$ the energy in m.f. will be
\be E_\pi (eB) = \sqrt{ |eB|+\bar m^2},\label{75}\ee where $\bar m^2$ can
depend on m.f. more slowly than $|eB|$.

This kind of  behavior was found on the lattice \cite{20}, and we shall find
below whether  it appears in our formalism and what $\bar m^2$ is.

Actually,  the behavior $m_{\pi^+} (eB)$ in  (\ref{75}), found on the lattice,
shows that $\bar m^2 \cong m^2_\pi (0)$ and  $\pi^+$ at $eB> m^2_\pi(0)$ can be
considered to some extent as an elementary pseudoscalar meson  seemingly
without internal $q\bar q$ structure. However, the derivation of the GMOR
relation for $\pi^+$ meson similarly to the $\pi^0$ case does not work for two
reasons. First of all,the cancellation in the $\Delta \bar N_{33}(0)$ term
which we observed for $\pi^0$, in the case of $\pi^+$ is absent. Secondly, the
total charge motion of $\pi^+$ in m.f. creates its own quantum energy $\Delta E
\sim eB$ which adds to $m^2_\pi$, as it is seen e.g. in (\ref{75}). Hence, the
GMOR relations do not apply to $\pi^+$ at $eB\ga m^2_\pi$ and $\pi^+$ mass
$m_{\pi^+} (eB)$ does not vanish in the limit $m_q, m_{\bar q}\to 0$. We shall
show below, however, that the behavior $m_{\pi^+}(eB)$ at $eB\ga\sigma$ can
display the $q\bar q$ structure and, moreover, the lattice data \cite{20}
possibly show the beginning of the new pattern.

We start with the expressions (\ref{30}), (\ref{31}) for $\rho^+ (S_z=0)$ and
$\pi^+$ states, which can be expressed as combinations $\frac{1}{\sqrt{2}}
(|+->\pm |-+>)$ of $(u\bar d)$ spin projected states. These two states can be
considered first in the approximation $eB\gg \sigma$, when $u$ and $\bar d$
quarks are independent, then \be M_{+-}(B) = \left (\sqrt{m^2_u+p^2_z}+
\sqrt{m^2_d+ p^2_z+ 2|e_d|B}\right)_{P_z =0}\approx \sqrt{\frac23
eB}\label{76}\ee

\be M_{-+}(B) = \left(\sqrt{m^2_u+p^2_z+2e_u B}+ \sqrt{m^2_d+
p^2_z}\right)_{P_z =0}\approx \sqrt{\frac43 eB}.\label{77}\ee

These two curves $M_{+-}(B)$ and $M_{-+}(B)$ are below and above the
``elementary'' behavior  $\sqrt{m^2_\pi+eB},$ Eq. (\ref{75}), see Fig. 2.

However, we have not taken into account the $hf$ interaction, which mixes these
two states,  and therefore one should diagonalize the spin-dependent part of
interaction \be M\cong \frac{eB}{3\omega_1} + \frac{eB}{6\omega_2} +
\frac{\omega_1+\omega_2}{2} + V_{SD}\label{78}\ee (see a similar treatment of
the neutral meson in \cite{14}) \be V_{SD} = a \vesig_1\vesig_2 +
\frac{eB\sigma_{1z}}{3\omega_1} -\frac{eB\sigma_{2z}}{6\omega_2}\label{79}\ee

As it  is seen from (\ref{78}),(\ref{79}) the  stationary values of
$\omega_1,\omega_2$ denoted as $\omega_1^{(0)}, \omega_2^{(0)}$ depend on the
state, and at $eB\gg \sigma$ \be \omega_1^{(0)} (+-) \cong \bar
m,~~\omega_2^{(0)} (+-) \cong \sqrt{\frac23 eB}, ~~\omega_1^{(0)} (-+) \cong
\sqrt{\frac43 eB}, ~~\omega_2^{(0)} (-+) \cong \bar m,~~\bar m\approx
\sqrt{\sigma}.\label{80}\ee

From \cite{22} \be a=\frac{c}{\pi^{3/2} \sqrt{\lambda^2 + r^2_0}
 (\lambda^2+r^2_\bot)}, ~~ c=
 \frac{8\pi\alpha_s}{9\omega_1\omega_2}\label{81}\ee
 and $\lambda\sim 1$ GeV$^{-1}$, while $r_0 \approx O(1/\sqrt{\sigma}), ~~
 r^2_\bot \sim \frac{2}{eB}.$

 Hence  the magnetic moment part of $V_{SD}$ (the last two terms on the r.h.s. of (\ref{79}))
  is always dominating for
 $  eB \ga \sigma$   and one expects  in this region that
 the asymptotic result for
 $\rho^+$ and $\pi^+$ are \be
 m_{as} (\rho^+ (S_z=0)) = M_{-+} (B) \approx \sqrt{\frac43 eB}\label{82}\ee

\be
 m_{as} (\pi^+ ) = M_{+-} (B) \approx \sqrt{\frac23 eB}.\label{83}\ee

 At smaller m.f., when $eB < \bar m^2 \approx \sigma$, one can diagonalize $V_{SD}$, and this procedure is given in Appendix. 

 The results of numerical calculations of asymptotic behavior for the $\pi^+$ and $\rho^+$ masses with the account of Coulomb and self-energy corrections are given in Fig. 2 (left graph). The contribution of spin-spin interaction can be neglected in this region. We extrapolate these asymptotic to small fields and compare them with the lattice data \cite{Bali} (right graph). One can see, that at large $eB>0.2$ GeV$^2$ the lattice data for $\pi^+$ possibly  prefer the lower asymptotic (\ref{83}),
rather than the ``elementary $\pi^+$ pion curve'' of Eq. (\ref{75}).

\begin{figure}
  \center{
  \includegraphics[height=5.5cm]{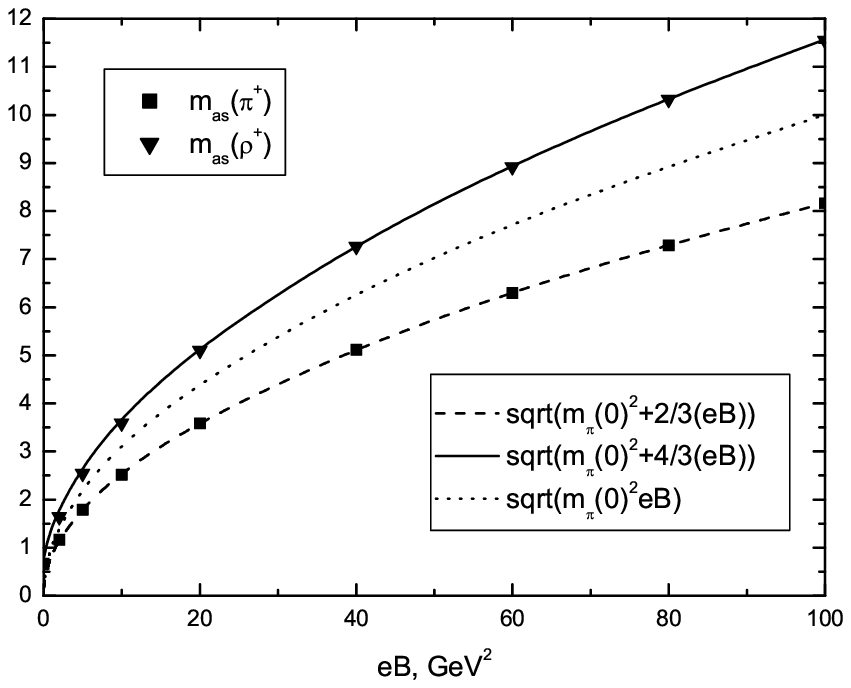}
  \includegraphics[height=5.5cm]{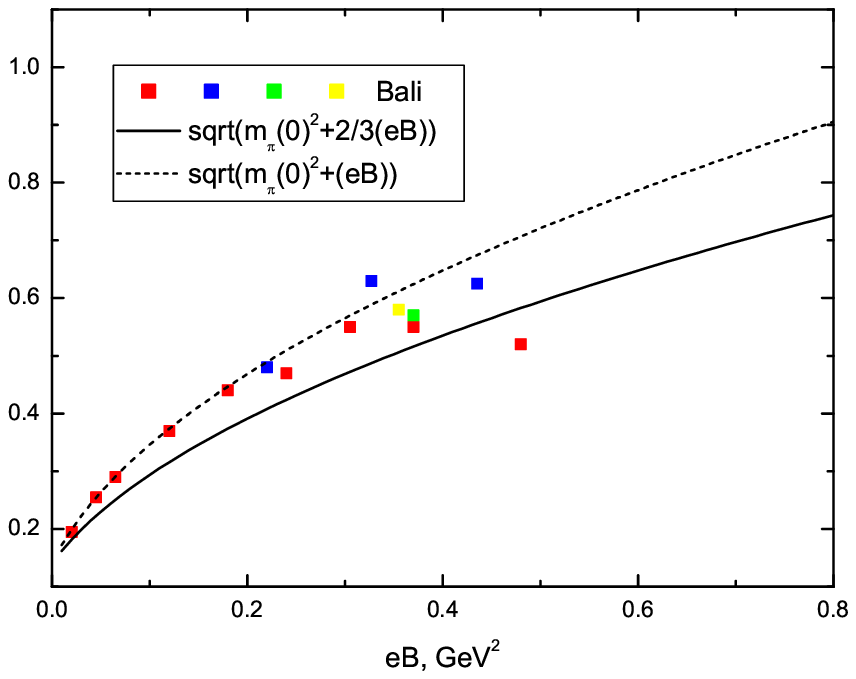}
    \caption{The masses of charged $\pi^+$ and $\rho^+$ mesons (in GeV) as a function of magnetic field strength at asymptotically large fields (left) and in the region $eB<1$ GeV$^2$ in comparison with lattice data of \cite{Bali} (right). \label{fig2}}}
\end{figure}

\section{Discussion of results and comparison to lattice data  and chiral
perturbation theory}

As was discussed above, our two examples,  $\pi^0$ and $\pi^+(\pi^-)$ mesons
behave quite differently at strong m.f. and while the first  obeys GMOR
relations, the charged meson looses all chiral properties at $eB > 0.1$
GeV$^2$. These facts are in  agreement with the results of chiral perturbation
theory \cite{23}-\cite{28}. In particular, it was shown in \cite{25,26} that
GMOR relations hold for the $\pi^0$ meson, while they are violated for the
$\pi^+$, and $\pi^0$  retains its NG properties in chiral perturbation theory.

However, as shown in \cite{16} and above, both $\lan \bar q q\ran$ and
$f^2_\pi$ are not any more objects of ChPT in strong m.f. and at $eB> m^2_\pi$
the $q\bar q$ degrees of freedom define the values of $\lan \bar q q\ran$ and
$f^2_\pi$.

This in particular is present in the m.f. dependence of $m^2_{\pi^0}$, which
according to ChPT is \cite{25, 26} \be \frac{m^2_{\pi^0}(eB)}{m^2_{\pi^0}(0)}=1 -
\frac{eB}{16\pi^2} \ln 2, \label{90}\ee and $e$ is the meson charge in ChPT,
while in the $(q\bar q)$ system two components $(\bar u u)$ and $(\bar d d)$
enter in an admixture, with $e_q = \frac23 e$ or $\frac13 e$. We compare the
dependence (\ref{90}) with our result (\ref{74}) in Fig. 1.

For $\pi^+$ meson  the ChPT is not applicable for $eB > m^2_\pi$, while the
$q\bar q$ structure is clearly seen for $eB> \sigma $,  as it is clear from
Fig.2, where the curve $m^2_{\pi^+} (eB)$ deflects from $m^2_{\rho^+}(eB)$, as
discussed in the previous section.

As it is, one can distinguish three regions: 1) $eB\la m^2_\pi(0)$, 2) $
m^2_\pi(0)\ll eB\la\sigma$, 3) $eB\gg\sigma$,  where different dominant
mechanisms of meson mass formation are present. In the region 1) the ChPT is
active for NG mesons, while in the region 2) the  $q\bar q$ structure is
evident and both m.f. effects and strong $q\bar q$ interaction (confinement and
gluon exchange) are important. Finally in the region 3) one can consider $q$
and $\bar q$ in $\pi^+$ as independent in the strong m.f. with asymptotic
calculated in section 4, while for $\pi^0$ the situation is more complicated
and the mass is defined by GMOR relations with $\lan q \bar q\ran$ and
$f^2_\pi$ computed in the nonchiral theory.

The authors are grateful  to N.~O.~Agasian,  M.~A.~Andreichikov and B.~O.~Kerbikov
for  useful discussions.

\newpage
\vspace{2cm}
 \setcounter{equation}{0}
\renewcommand{\theequation}{A \arabic{equation}}

\hfill {\it  Appendix   }


 \vspace{1cm}

\setcounter{equation}{0} \def\theequation{A.  \arabic{equation}}

 \be
 m(\rho^+ (S_z=0),B )= \frac12 (M_{11} + M_{22}) + \sqrt{ \left( \frac{M_{11} -
 M_{22}}{2}\right)^2 + 4 a_{12} a_{21}},\label{84}\ee

\be
 m(\pi^+, B)  = \frac12 (M_{11} + M_{22}) - \sqrt{ \left( \frac{M_{11} -
 M_{22}}{2}\right)^2 + 4 a_{12} a_{21}},\label{85}\ee
 where \be M_{11} = \bar M - \frac{eB}{3\omega_1(+-)} +
 \frac{eB}{6\omega_1(+-)}-a_{11} (+-)\label{86}\ee

\be M_{22} = \bar M +\frac{eB}{3\omega_1(-+)} -
 \frac{eB}{6\omega_1(-+)}-a_{22} (-+)\label{87}\ee
and $a_{ik}$ is given in (\ref{81}), with $c_{ik}$ defined as

\be c_{11}= \frac{8\pi\alpha_s}{9 \omega_1 (+-) \omega_2 (+-)}, ~~c_{22}=
\frac{8\pi\alpha_s}{9 \omega_1 (-+) \omega_2 (-+)},\label{88}\ee and \be
c_{12}c_{21} = \left( \frac{8\pi\alpha_s}{9}\right)^2
\frac{1}{\omega_1(+-)\omega_1(-+) \omega_2(+-) \omega_2(-+)}.\label{89}\ee

The values of $\bar M$ can be calculated from (\ref{30}) or (\ref{32}), or else for $eB\gg \sigma$ they can be
estimated  as $\bar M \approx \omega_1 + \omega_2 = \sqrt{e_1 B} + \sqrt{e_2
B}$.

\end{document}